\documentclass[aps,prl,twocolumn,superscriptaddress,amsmath,amssymb,floatfix,longbibliography]{revtex4-2}

\usepackage{graphicx}
\usepackage{mathtools}
\usepackage{amsthm}
\usepackage{bm}
\usepackage{pifont}
\usepackage[shortlabels]{enumitem}
\usepackage{xcolor}

\usepackage[colorlinks=true,linkcolor=blue!60!black,citecolor=blue!60!black,urlcolor=blue!60!black]{hyperref}

\usepackage{tikz}
\usetikzlibrary{decorations.pathreplacing}
\usepackage{graphicx}
\usepackage[colorlinks=true,linkcolor=blue!60!black,citecolor=blue!60!black,urlcolor=blue!60!black]{hyperref}

\definecolor{revisiongreen}{RGB}{0,115,70}

\newcommand{\CPTP}{\operatorname{CPTP}}
\newcommand{\Tr}{\operatorname{Tr}}
\newcommand{\id}{\operatorname{id}}
\newcommand{\cc}{\mathrm{cc}}
\newcommand{\rmarg}{R_{\mathrm m}^{\mathcal O}}
\newcommand{\cO}{\mathcal O}
\newcommand{\cF}{\mathcal F}

\newcommand{\Rm}{R_{\mathrm m}}
\newcommand{\norm}[1]{\left\lVert #1\right\rVert_1}
\newcommand{\ketbra}[1]{|#1\rangle\!\langle#1|}
\newcommand{\ket}[1]{|#1\rangle}
\newcommand{\bra}[1]{\langle#1|}
\newcommand{\Rc}{R_{\mathrm c}}
\newcommand{\Rmc}{R_{\mathrm{mc}}}
\newcommand{\Dm}{D_{\mathrm m}}
\newcommand{\Cm}{C_{\mathrm m}}
\newcommand{\Ed}{E_{\mathrm d}}
\newcommand{\ccto}{\xrightarrow{\;\mathrm{cc}\;}}
\newcommand{\cmark}{\ding{51}}
\newcommand{\xmark}{\ding{55}}

\theoremstyle{definition}
\newtheorem{definition}{Definition}
\newtheorem{assumption}{Assumption}

\theoremstyle{plain}
\newtheorem{proposition}{Proposition}
\newtheorem{lemma}{Lemma}
\newtheorem{theorem}{Theorem}
\newtheorem{corollary}{Corollary}

\begin{document}

\title{Equivalence of Correlated Catalytic and Marginal Asymptotic Transformations\\ in Quantum Resource Theories}

\author{Jingsong Ao}
\affiliation{Institute of Fundamental Technological Research, Polish Academy of Sciences, Pawi\'nskiego 5B, 02-106 Warsaw, Poland}
\author{Aby Philip}
\affiliation{Institute of Fundamental Technological Research, Polish Academy of Sciences, Pawi\'nskiego 5B, 02-106 Warsaw, Poland}
\author{Alexander Streltsov}
\affiliation{Institute of Fundamental Technological Research, Polish Academy of Sciences, Pawi\'nskiego 5B, 02-106 Warsaw, Poland}

\date{\today}

\begin{abstract} 
We prove that correlated-catalytic state conversion is possible if and only if the marginal asymptotic conversion rate is at least one. Under general operational assumptions, this equivalence holds for arbitrary states in broad classes of entanglement, coherence, and magic theories. It reduces the characterization of correlated-catalytic convertibility to the computation of marginal asymptotic rates, and yields transitivity and equality of catalytic and marginal conversion rates. Under local operations and classical communication, we fully characterize correlated-catalytic conversion from arbitrary bipartite states to entangled pure states in terms of distillable entanglement. We show that correlated catalysts cannot increase the distillable entanglement or distillable secret key of any bipartite state, and in particular cannot make bound entanglement distillable. Extending our construction to adaptive communication protocols, we further prove that correlated catalysts cannot enhance the two-way private-state generation capacity of any finite-dimensional quantum channel.
\end{abstract}

\maketitle
Quantum resource theories (QRTs) provide a general framework for characterizing state transformations under a restricted class of operations~\cite{chitambar2019}. Prominent examples include entanglement~\cite{horodecki2009}, coherence~\cite{baumgratz2014,streltsov2017rmp}, asymmetry~\cite{bartlett2007,gour2008,marvian2016}, magic~\cite{veitch2014}, and athermality~\cite{brandao2015pnas}. In many resource theories, catalysis can enlarge the set of achievable transformations: an auxiliary system, the catalyst, enables a conversion $\rho\to\sigma$ while being returned in its initial state~\cite{jonathan1999,datta2023,lipkabartosik2024}. However, allowing the catalyst to be returned only approximately can trivialize state convertibility in certain settings, as embezzling catalysts can enable essentially arbitrary transformations~\cite{vandam2003,lipkabartosik2024}. In \emph{correlated} catalysis, the catalyst is instead returned exactly in its marginal state but may become correlated with the system~\cite{mueller2018,shiraishi2021,kondra2021,wilming2021,rubboli2022}. This relaxation can substantially enhance state convertibility. In particular, state convertibility can be characterized by a single monotone, such as the free energy in thermodynamic settings~\cite{mueller2018,shiraishi2021,shiraishi2025} and the entanglement entropy for pure bipartite states~\cite{kondra2021}. Moreover, correlated catalysis can even enable unbounded amplification of coherence~\cite{ding2021,takagi2022,shiraishi2024,kondra2024}.

Correlated catalysis is closely tied to asymptotic transformations of many copies~\cite{duan2005,kondra2021,takagi2022}, in particular to those in which only the single-copy marginals of the output have to approximate the target~\cite{ganardi2024,shiraishi2024}. A cyclic catalyst built from many copies turns such transformations at rate one into correlated-catalytic ones; the construction goes back to Ref.~\cite{duan2005}, and Ref.~\cite{takagi2022} established it in general QRTs for standard asymptotic transformations. The converse is far less understood. In entanglement theory, it was proven for distillable states and, under operations preserving the positivity of the partial transpose (PPT), for all states~\cite{ganardi2024}. Moreover, correlated catalysis can neither distill entanglement from PPT states~\cite{lami2024} nor render entanglement manipulation reversible~\cite{lami2024,ao2026ppt}. For local operations and classical communication (LOCC), however, it has remained open whether correlated catalysis can make states with non-positive partial transpose (NPT) but zero distillable entanglement distillable, should such states exist~\cite{divincenzo2000,dur2000,ganardi2024}. More generally, it is unclear which structural features of a QRT make correlated catalysis and marginal asymptotics coincide.

In this Letter, we identify such features. We show that in every QRT admitting classical control and processing, permutations of identical systems, full-rank free states, and free informationally complete (IC) measurements as free operations, correlated catalysis converts $\rho$ into $\sigma$ if and only if the marginal asymptotic rate is at least one. 
The equivalence holds for arbitrary input and target states in a broad range of resource theories, including entanglement, coherence, and magic. Through correlated catalysts, it connects one-shot state conversion in the microscopic regime with marginal asymptotic conversion in the asymptotic limit of resource manipulation. Earlier constructions showed that correlated catalysts can implement asymptotically achievable transformations~\cite{duan2005,takagi2022,ganardi2024}. We establish the converse: every one-shot correlated-catalytic conversion can be achieved at unit marginal asymptotic rate. Thus, although catalysts can enable otherwise impossible one-shot transformations, their power does not exceed that of marginal asymptotic conversion. For LOCC, this removes the distillability assumption of Ref.~\cite{ganardi2024} and proves that correlated catalysis cannot increase the distillable entanglement of any state. We then use the results regarding correlated catalyst and marginal conversions rates to show that correlated catalyst cannot provide any improvement in the rate of secret key distillation from bipartite states nor can it increase the two-way private-state generation capacity of any finite-dimensional quantum channels.

Outside the scope of our assumptions, correlated catalysis may be strictly more powerful than marginal asymptotic conversion. Relevant settings include resource theories of asymmetry associated with certain symmetry groups~\cite{marvian2013} and the resource theory of nonlocality based on local operations and shared randomness (LOSR)~\cite{schmid2020losr,schmid2023losr}. This raises the broader question of which physical restrictions allow correlated catalysts to outperform marginal asymptotic conversion. 

\textit{Framework.}---We now introduce the resource-theoretic framework and the assumptions under which our main theorem holds.

A QRT specifies sets $\cO(A\to B)$ of free CPTP maps between finite-dimensional systems, containing identities and closed under composition, with $\cF(A):=\cO(\mathbf 1\to A)\neq\varnothing$ and $\cO(A\to\mathbf 1)=\{\Tr_A\}$, where $\mathbf 1$ is the trivial system~\cite{chitambar2019}.

We assume tensor-product closure: $\Lambda\otimes\Phi\in\cO(AC\to BD)$ whenever $\Lambda\in\cO(A\to B)$ and $\Phi\in\cO(C\to D)$. Thus discarding subsystems and appending free states are free. Classical registers have distinguished bases $\{\ket x\}$. We make the following assumptions:

\begin{enumerate}[label=(A\arabic*),leftmargin=2.6em, itemsep=1pt,topsep=2pt]
\item \emph{Classical control:}
if $\Lambda_x\in\cO(A\to B)$ for all $x$, then
\[
\xi^{AX}\mapsto
\sum_x\Lambda_x(\bra x\xi^{AX}\ket x)\otimes\ketbra x^X
\]
belongs to $\cO(AX\to BX)$.
\item \emph{Classical processing:} every stochastic map $\ketbra x\mapsto\sum_y p(y|x)\ketbra y$ is free.
\item \emph{Permutations:} permutations of identical subsystems is free.
\end{enumerate}

\begin{enumerate}[label=(A\arabic*),start=4,leftmargin=2.6em,itemsep=1pt,topsep=2pt]
\item \emph{Full-rank free states:}
every system $C$ admits $\omega^C\in\cF(C)$ with $\omega^C>0$.
\item \emph{Free IC measurements:}
every system $B$ admits a finite POVM $\{M_z\}$ spanning $\operatorname{Herm}(B)$, such that the following measurement channel is free:
\[
\xi\mapsto\sum_z\Tr(M_z\xi)\ketbra z
\]
\end{enumerate}

For correlated catalytic conversion, we write $\rho^A\ccto\sigma^B$ if for every $\varepsilon>0$ there are a finite-dimensional catalyst $\tau^C$ and $\Lambda\in\cO(AC\to BC)$ such that $\mu^{BC}:=\Lambda(\rho^A\otimes\tau^C)$ satisfies $\mu^C=\tau^C$ and $\norm{\mu^B-\sigma}<\varepsilon$~\cite{ganardi2024}. The \emph{marginal asymptotic rate} $\Rm(\rho\to\sigma)$ is the supremum of all $r\ge0$ such that for all $\varepsilon,\delta>0$ there exist $n,m\in\mathbb N$ and $\Lambda\in\cO(A^{\otimes n}\to B^{\otimes m})$ with
\begin{equation}
    \max_{1\le i\le m}\norm{[\Lambda(\rho^{\otimes n})]^{B_i}-\sigma}<\varepsilon,
    \qquad \frac mn>r-\delta,
    \label{eq:Rm}
\end{equation}
where $B_i$ denotes the $i$th output~\cite{ganardi2024}. In contrast to the standard asymptotic rate $R(\rho\to\sigma)$, which requires $\Lambda(\rho^{\otimes n})\approx\sigma^{\otimes m}$, only the marginals are constrained; hence $R\le\Rm$.

\textit{Main result.}--- Our central result is the following.

\begin{theorem}\label{thm:main}
In every QRT satisfying \textup{(A1)--(A5)},
\begin{equation}
    \rho\ccto\sigma\quad\Longleftrightarrow\quad \Rm(\rho\to\sigma)\ge1 .
    \label{eq:main}
\end{equation}
\end{theorem}

The implication $\Rm(\rho\to\sigma)\ge1\Rightarrow\rho\ccto\sigma$ follows from the known cyclic construction~\cite{takagi2022,ganardi2024,shiraishi2024,lipkabartosik2021}. For completeness, we include a proof under assumptions \textup{(A1)--(A3)} in the Supplemental Material (SM)~\cite{SM}. We now turn to the converse, the main contribution of this work.

\begin{figure}[t]
\centering
\includegraphics[width=\columnwidth]{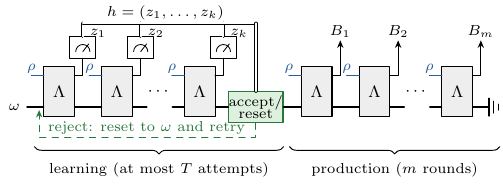}
\caption{Protocol of Proposition~\ref{prop:learning}. Starting from a full-rank free memory $\omega\ge c\tau$, each of $k$ rounds applies the catalytic protocol $\Lambda$ to a fresh copy of $\rho$ and the memory, and measures the system output with a free IC POVM. The resulting history $h$ is accepted with probability $cQ(h)/P(h)$, where $P$ and $Q$ are the history distributions for initial memories $\omega$ and $\tau$. Acceptance changes the history weights from $P$ to $Q$ while leaving each conditional memory state unchanged. On rejection the memory is reset to $\omega$, with at most $T$ attempts. The subsequent $m$ production rounds retain the outputs $B_1,\dots,B_m$, whose marginals are all close to $\sigma$.}
\label{fig:protocols}
\end{figure}

\begin{proposition}\label{prop:learning}
Under assumptions \textup{(A1)}, \textup{(A2)}, \textup{(A4)}, and \textup{(A5)}, $\rho\ccto\sigma$ implies $\Rm(\rho\to\sigma)\ge1$.
\end{proposition}
The full proof can be found in Supplemental Material (SM)~\cite{SM}. We provide a brief proof sketch below.

\begin{proof}[Proof sketch]
Fix a catalytic protocol $\Lambda$ with catalyst $\tau^C$ and output $\mu^{BC}=\Lambda(\rho^A\otimes\tau^C)$, where $\mu^C=\tau^C$ and $\|\mu^B-\sigma^B\|_1<\varepsilon$. Choose a full-rank free state $\omega^C\ge c\tau^C$ with $0<c<1$.
The protocol is illustrated in Fig.~\ref{fig:protocols}.

Each round feeds a fresh copy of $\rho$ in the system and the memory in $C$ into $\Lambda$, measures the system output with a free IC measurement, and passes the memory to the next round.
After $k$ rounds, the initial memories $\omega$ and $\tau$ give the respective ensembles
\[
    \{P(h),\omega_h\}_h,
    \qquad
    \{Q(h),\tau_h\}_h,
\]
where $h$ is the measurement history. 
We have $P(h)\ge cQ(h)$, while exact catalyst return gives $\sum_hQ(h)\tau_h=\tau$.

Starting from $\omega$, after observing $h$ we toss a coin with heads probability $cQ(h)/P(h)$.
On heads we keep $\omega_h$; on tails we reset to $\omega$.
The resulting ensemble is
\[
    \bigl\{(cQ(h),\omega_h)\bigr\}_h
    \;\cup\;
    \bigl\{(1-c,\omega)\bigr\}.
\]
Thus, conditioned on heads, the ensemble is $\{Q(h),\omega_h\}_h$: the weights have changed from $P$ to $Q$, but the memory state associated with each history is unchanged.

Repeat at most $T$ times, keeping the memory from the first successful attempt. With $f_T=(1-c)^T$, the resulting ensemble is
\[
    \bigl\{((1-f_T)Q(h),\omega_h)\bigr\}_h
    \;\cup\;
    \bigl\{(f_T,\omega)\bigr\}.
\]
For each branch of the ensemble, we take the corresponding state from the ensemble above as the initial memory state in $C$ and run $m$ further rounds. In each round, a fresh copy of $\rho$ and the current memory in $C$ are fed into $\Lambda$. The system output is retained without measurement, while the updated memory is passed to the next round. The resulting $m$ system outputs are collected as the final output.

The key is to compare the future system outputs generated by the two ensembles $\{Q(h),\omega_h\}_h$ and $\{Q(h),\tau_h\}_h$. Since the measurement is informationally complete, the trace distance between quantum states can be bounded by a constant times the distance between their measurement outcome distributions. It therefore suffices to compare the corresponding classical probabilities. Conditioned on a history $h$, these are the predictions $P(\cdot\mid h)$ and $Q(\cdot\mid h)$ for future measurement outcomes. Since $P(h)\ge cQ(h)$ at every history length, the Blackwell--Dubins theorem~\cite{BlackwellDubins1962} guarantees that these predictions become arbitrarily close, on average over $Q$-distributed histories and uniformly over all future lengths. Informational completeness then implies that the output marginals of the two ensembles are arbitrarily close, uniformly over all future rounds. The reference ensemble has average memory $\tau$ and therefore produces $\mu^B$ in every round. Finally, the failure branch contributes an arbitrarily small error for sufficiently large $T$.

Consequently, $k$ and $T$ can be chosen so that every output marginal is arbitrarily close to $\sigma$, independently of $m$.
The protocol consumes at most $Tk+m$ input copies, giving
\[
    \frac{m}{Tk+m}\longrightarrow1.
\]
Hence $\Rm(\rho\to\sigma)\ge1$.
Full details are given in the supplemental material~\cite{SM}.
\end{proof}

The proof admits a natural interpretation in terms of predictive learning. Rather than reconstructing the catalyst, we need only reproduce its predictions for future outputs. A full-rank free memory induces a probability law $P$ that dominates the law $Q$ associated with the catalyst. By the Blackwell--Dubins merging-of-opinions theorem~\cite{BlackwellDubins1962}, their conditional predictions become arbitrarily close, on average over $Q$-distributed histories and uniformly over all future lengths. Crucially, informational completeness (A5) converts this agreement of classical predictions into closeness of quantum output marginals, even though the underlying memory states need not approach each other. Thus the catalytic effect can be reproduced without reconstructing the catalyst itself.

\textit{Consequences.}---Theorem~\ref{thm:main}  establishes the
transitivity of correlated-catalytic convertibility, even
though correlations with the catalyst obstruct the direct
composition of catalytic protocols.

\begin{corollary}[Transitivity]\label{cor:transitivity}
Under \textup{(A1)--(A5)},
\begin{equation}
    \rho\ccto\sigma\ \text{ and }\ \sigma\ccto\chi
    \quad\Longrightarrow\quad\rho\ccto\chi .
    \label{eq:transitivity}
\end{equation}
\end{corollary}

Theorem~\ref{thm:main}  also implies that the optimal correlated catalytic rates coincide with the marginal asymptotic rate. Let $\Rc$ and $\Rmc$ denote the optimal rates with exact catalyst return, requiring joint and marginal
approximation of the target outputs, respectively.

\begin{corollary}[Catalytic rates]\label{cor:rates}
Under \textup{(A1)--(A5)},
\begin{equation}
    \Rc(\rho\to\sigma)=\Rmc(\rho\to\sigma)
    =\Rm(\rho\to\sigma).
    \label{eq:rates}
\end{equation}
\end{corollary}

Both results are proved in the Supplemental Material~\cite{SM}.

\textit{Scope of the assumptions.}---
Table~\ref{tab:theories} summarizes the validity of the assumptions \textup{(A1)--(A5)} for various resource theories of interest. For theories satisfying all the assumptions, correlated-catalytic convertibility is completely characterized by the marginal asymptotic rate $\Rm$, although computing this rate is generally difficult. 

The two directions within Theorem~\ref{thm:main} nevertheless have different scopes. The cyclic construction uses tensor-product closure, \textup{(A3)}, and only restricted forms of \textup{(A1)} and \textup{(A2)}, available in all the listed theories. The converse proof instead requires tensor-product closure and \textup{(A1)}, \textup{(A2)}, \textup{(A4)}, and \textup{(A5)}.

For resource theories that do not satisfy all the assumptions, correlated catalysis may be strictly more powerful than marginal asymptotic conversion. Such a separation occurs in asymmetry theories for finite groups: exact catalysis enables arbitrary state transformations~\cite{marvian2013}, whereas marginal asymptotic conversion cannot generate asymmetry from symmetric inputs. 

\begin{table*}[t]
\caption{\label{tab:theories}
Tensor-product closure (TP) and assumptions (A1)--(A5)
for representative resource theories.
Here \cmark\ denotes satisfaction, \xmark\ violation,
and $\circ$ a restricted form sufficient for the
forward cyclic construction but not for the converse proof.}
\small
\setlength{\tabcolsep}{4pt}
\renewcommand{\arraystretch}{1.12}
\begin{ruledtabular}
\begin{tabular}{@{}llcccccc@{}}
Resource & Free operations
 & TP & A1 & A2 & A3 & A4 & A5\\
\hline
Entanglement
 & Local operations and classical communication~\cite{horodecki2009}
 & \cmark & \cmark & \cmark & \cmark & \cmark & \cmark\\
 & Separable operations~\cite{chitambar2019}
 & \cmark & \cmark & \cmark & \cmark & \cmark & \cmark\\
 & Positive-partial-transpose-preserving operations~\cite{rains2001}
 & \cmark & \cmark & \cmark & \cmark & \cmark & \cmark\\
\hline
Coherence
 & Maximally incoherent operations~\cite{chitambar2016}
 & \cmark & \cmark & \cmark & \cmark & \cmark & \cmark\\
 & Incoherent operations~\cite{baumgratz2014}
 & \cmark & \cmark & \cmark & \cmark & \cmark & \cmark\\
 & Dephasing-covariant incoherent operations~\cite{chitambar2016}
 & \cmark & \cmark & \cmark & \cmark & \cmark & \xmark\\
 & Strictly incoherent operations~\cite{chitambar2016}
 & \cmark & \cmark & \cmark & \cmark & \cmark & \xmark\\
\hline
Magic
 & Stabilizer operations~\cite{veitch2014}
 & \cmark & \cmark & \cmark & \cmark & \cmark & \cmark\\
 & Completely stabilizer-preserving operations~\cite{seddon2019}
 & \cmark & \cmark & \cmark & \cmark & \cmark & \cmark\\
 & Completely positive-Wigner-preserving operations~\cite{wang2019}
 & \cmark & \cmark & \cmark & \cmark & \cmark & \cmark\\
\hline
Asymmetry
 & Group-covariant operations~\cite{gour2008}
 & \cmark & \cmark & \cmark & \cmark & \cmark & \xmark\\
\hline
Athermality
 & Gibbs-preserving operations~\cite{faist2015}
 & \cmark & \cmark & $\circ$ & \cmark & \cmark & \cmark\\
 & Covariant Gibbs-preserving operations~\cite{shiraishi2025}
 & \cmark & \cmark & $\circ$ & \cmark & \cmark & \xmark\\
 & Thermal operations~\cite{brandao2015pnas}
 & \cmark & \cmark & $\circ$ & \cmark & \cmark & \xmark\\
\hline
Nonlocality
 & Local operations and shared randomness~\cite{schmid2020losr}
 & \cmark & $\circ$ & $\circ$ & \cmark & \cmark & \cmark
\end{tabular}
\end{ruledtabular}
\end{table*}

\textit{Applications.}---
Theorem~\ref{thm:main}  reduces the characterization of
correlated-catalytic convertibility to computing
the marginal asymptotic rate $\Rm$.
Although this rate is generally difficult to evaluate,
explicit formulas are available in some important cases.

We begin by looking into entanglement theory.
For bipartite LOCC, Theorem~\ref{thm:main} establishes a limitation
on correlated catalysis without assuming that the input
state is distillable. For any bipartite state $\rho$ and
entangled pure state $\phi$, the marginal asymptotic rate
is known to satisfy~\cite{ganardi2024}
\begin{equation}
\Rm(\rho\to\phi)
    =R(\rho\to\phi)
    =\frac{\Ed(\rho)}{S(\phi_A)},
\end{equation}
where $\Ed$ denotes distillable entanglement and $S$
is the von Neumann entropy.
Combining this with Theorem~\ref{thm:main}  and the composition
of marginal rates, we obtain the following.

\begin{corollary}[Entanglement]\label{cor:entanglement}
Under LOCC, for arbitrary bipartite states $\rho,\sigma$
and any entangled pure state $\phi$,
\begin{align}
    \rho\ccto \sigma
    &\implies \Ed(\rho)\ge\Ed(\sigma),\\
    \rho\ccto\phi
    &\iff \Ed(\rho)\ge S(\phi_A).
\end{align}
\end{corollary}

Thus, correlated catalysis cannot increase distillable
entanglement, even for undistillable input states.
In particular, it cannot make any bound-entangled state
distillable, including hypothetical NPT bound-entangled
states~\cite{divincenzo2000,dur2000}.
For LOCC, this removes the distillability assumption
of Ref.~\cite{ganardi2024} and extends the PPT no-go
result of Ref.~\cite{lami2024}.

We now look into secret key distillation and private communication.
Theorem~\ref{thm:main} also has implications for quantum cryptography. The distillable secret key $K_D(\rho)$ is the optimal asymptotic rate of secret bits extracted from $\rho$ by LOCC, secure against an eavesdropper holding a purification and the public communication record ~\cite{devetak2005,horodecki2009private}.
Unlike distillable entanglement, $K_D$ can be positive even for PPT bound-entangled states~\cite{horodecki2005key}. Nevertheless, Theorem~\ref{thm:main} implies that
\begin{equation}
    \rho\ccto\sigma
    \quad\Longrightarrow\quad
    K_D(\rho)\ge K_D(\sigma).
    \label{eq:key-monotonicity}
\end{equation}

A similar limitation extends to quantum communication. For a finite-dimensional quantum channel $\mathcal N$, let $P_2(\mathcal N)$ denote its private-state generation capacity with unlimited two-way classical communication, which equals its two-way secret-key capacity ~\cite{horodecki2009private,pirandola2017}. Let $P_{2,\mathrm{cc}}(\mathcal N)$ denote the corresponding private-state generation capacity assisted by a finite-dimensional bipartite catalyst whose joint state is returned exactly, possibly correlated with the output. Applying our construction to adaptive communication protocols, we obtain \begin{equation}     P_{2,\mathrm{cc}}(\mathcal N)=P_2(\mathcal N).     \label{eq:catalytic-private-capacity} \end{equation} Thus correlated catalysis can neither increase the distillable secret key of a state nor enhance the two-way private-state generation capacity of a finite-dimensional channel. Formal definitions and proofs are given in the supplemental material~\cite{SM}.

\textit{Discussion.}---In this work, we have shown that correlated-catalytic conversion is possible if and only if the marginal asymptotic rate is at least one, under operational assumptions satisfied by broad classes of entanglement, coherence, and magic theories.

This equivalence also implies that correlated catalysts cannot increase marginal asymptotic conversion rates. Under LOCC, our results fully characterize conversion from any bipartite state to entangled pure states in terms of distillable entanglement and prove that correlated catalysis cannot increase distillable entanglement or secret key for any bipartite state. Extending our construction to adaptive communication protocols further shows that correlated catalysts cannot enhance the two-way private-state generation capacity of any finite-dimensional quantum channel. 

Several questions remain open. When $\Rm(\rho\to\sigma)<1$, deterministic correlated-catalytic conversion is impossible under our assumptions. It is natural to ask whether this subunit rate instead characterizes the optimal success probability of a probabilistic correlated-catalytic conversion, and how the answer depends on the catalyst-return conditions in the success and failure branches. Such a connection would complement known tradeoffs between success probability and overhead in catalytic distillation~\cite{fang2024}. The computation of $\Rm$ also remains largely unexplored in magic and multipartite entanglement. If explicit rates for magic states were obtained, Theorem~\ref{thm:main} would immediately provide conversion criteria for magic states which would be relevant to quantum computation. In multipartite entanglement, graph states provide a natural starting point for investigating how entanglement structure constrains marginal conversion rates.

Finally, equality of the optimal conversion rates leaves open the complexity of the corresponding protocols. Catalytic and marginal asymptotic implementations may require different circuit depths and amounts of classical processing. Comparing how these costs scale with system size and target accuracy would help determine whether correlated catalysts can make a transformation more efficient to implement at a fixed conversion rate.

\textit{Acknowledgments.}---This work was supported by the National Science Centre Poland (Grant No. 2022/46/E/ST2/00115 and 2024/55/B/ST2/01590).

\emph{Use of AI tools.}— During the preparation of this manuscript, the authors used ChatGPT (OpenAI) to assist with language editing, literature searches, and discussions of mathematical arguments. The authors take full responsibility for the accuracy and integrity of all derivations, results, references, and conclusions presented in this work.

\bibliography{refs}

\clearpage
\onecolumngrid
\begin{center}
{\Large\bfseries Supplemental Material}\par\medskip

\end{center}

\newcommand{\freeccto}[1]{\xrightarrow[#1]{\cc}}
\section{Framework and assumptions}

\begin{definition}[Quantum resource theory]
For every pair of finite-dimensional physical systems $A$ and $B$, let
\[
    \mathcal O(A\to B)\subseteq \CPTP(A\to B)
\]
denote the set of free operations from $A$ to $B$. We assume:
\begin{enumerate}[label=(\roman*)]
    \item $\id_A\in\mathcal O(A\to A)$;
    \item if $\Lambda\in\mathcal O(A\to B)$ and
    $\Phi\in\mathcal O(B\to C)$, then
    $\Phi\circ\Lambda\in\mathcal O(A\to C)$;
    \item for every $A$,
    \[
        \mathcal O(\mathbf 1\to A)\neq\varnothing,
        \qquad
        \mathcal O(A\to\mathbf 1)\neq\varnothing
        ,
    \]
    where $\mathbf 1$ is the trivial system.
\end{enumerate}
The free states of $A$ are
\[
    \mathcal F(A):=\mathcal O(\mathbf 1\to A).
\]
\end{definition}
In particular, $\mathcal O(A\to\mathbf 1)=\CPTP(A\to\mathbf 1)=\{\Tr_A\}$.
\begin{definition}[Tensor-product closure]
The QRT is tensor-product closed if, for
$\Lambda\in\mathcal O(A\to B)$ and
$\Phi\in\mathcal O(C\to D)$,
\[
    \Lambda\otimes\Phi\in\mathcal O(AC\to BD).
\]
\end{definition}

\begin{assumption}[Classical operations and permutations]
\label{ass:prop1}
The following operations are free.

\begin{enumerate}[label=\textup{(A\arabic*)},leftmargin=3.2em]

\item \emph{Classical control.}
Let $X$ be a finite classical register with distinguished basis
$\{|x\rangle\}_{x\in X}$. For every family of free operations
\[
    \Lambda_x\in\mathcal O(A\to B),
    \qquad x\in X,
\]
the controlled channel
\[
    \omega^{AX}
    \longmapsto
    \sum_{x\in X}
    \Lambda_x\!\left(
        \langle x|\omega^{AX}|x\rangle_X
    \right)
    \otimes |x\rangle\!\langle x|^X
\]
belongs to
\[
    \mathcal O(AX\to BX).
\]

\item \emph{Classical processing.}
Let $X$ and $Y$ be finite classical registers. For every conditional probability distribution $p(y|x)$, the classical channel
\[
    |x\rangle\!\langle x|^X
    \longmapsto
    \sum_y p(y|x)|y\rangle\!\langle y|^Y
\]
belongs to $\mathcal O(X\to Y)$.

\item \emph{Permutations of identical subsystems.}
For every system $A$, every $n\in\mathbb N$, and every
permutation $\pi\in S_n$, the permutation channel
\[
    \mathcal P_\pi(\rho)
    :=
    U_\pi \rho U_\pi^\dagger
\]
belongs to
\[
    \mathcal O(A^{\otimes n}\to A^{\otimes n}).
\]

\end{enumerate}
\end{assumption}

\begin{definition}[Marginal asymptotic rate]\label{def:Rm}
For $\rho\in\mathcal D(A)$ and $\sigma\in\mathcal D(B)$, define
\begin{equation}
\rmarg(\rho\to\sigma)
:=
\sup\left\{
\begin{array}{@{}l|l@{}}
 r\ge0
 & \begin{array}{l}
 \forall\,\varepsilon,\delta>0,\ \exists\,n,m\in\mathbb N,\ \\[-1mm]
 \exists\,\Lambda\in\mathcal O(A^{\otimes n}\to B^{\otimes m})\ \text{such that}\\[1mm]
 \displaystyle
 \max_{1\le i\le m}
 \left\|
     \bigl[\Lambda(\rho^{\otimes n})\bigr]^{B_i}-\sigma
 \right\|_1<\varepsilon,
 \qquad
 \displaystyle \frac{m}{n}>r-\delta
 \end{array}
\end{array}
\right\}.
\label{eq:Rm-sm}
\end{equation}
\end{definition}

\begin{definition}[Correlated catalysis]\label{def:cc}
We write
\[
    \rho^A\xrightarrow[\mathcal O]{\cc}\sigma^B
\]
if, for every $\varepsilon>0$, there exist a finite-dimensional
catalyst $C$, a state $\tau^C$, and
$\Lambda\in\mathcal O(AC\to BC)$ such that, with
\[
    \mu^{BC}:=\Lambda(\rho^A\otimes\tau^C),
\]
one has
\[
    \mu^C=\tau^C,
    \qquad
    \|\mu^B-\sigma^B\|_1<\varepsilon.
\]
\end{definition}

\section{Cyclic construction}
\begin{proposition}[Cyclic construction]
\label{prop:cyclic}~\cite{takagi2022,ganardi2024,shiraishi2024,lipkabartosik2021}
In a QRT with tensor-product closure and assumptions
\textup{(A1)--(A3)},
\[
    R_m^{\mathcal O}(\rho\to\sigma)\ge1
    \quad\Longrightarrow\quad
    \rho\freeccto{\mathcal O}\sigma .
\]
\end{proposition}

\begin{proof}
By appending and discarding free states, we may without loss of generality assume that $\rho$ and $\sigma$ are states of the same system $S$.

Fix $\varepsilon,\delta>0$. Since
$\rmarg(\rho\to\sigma)\ge1$, Definition~\ref{def:Rm} gives integers $m\le n$ and a free operation $\Lambda\in\mathcal O(S^{\otimes n}\to S^{\otimes n})$
such that, with
\[
    \Gamma_n:=\Lambda(\rho^{\otimes n}),
\]
we have
\begin{equation}
    \|\Gamma_n^{S_j}-\sigma\|_1<\varepsilon,
    \qquad j=1,\ldots,m,
    \qquad
    \frac{m}{n}>1-\delta.
\label{eq:good}
\end{equation}
For $j=0,\ldots,n$, let
\[
    \Gamma_j:=\Tr_{S_{j+1}\cdots S_n}\Gamma_n,
    \qquad \Gamma_0:=1,
\]
and denote by
\[
    \Gamma_j^{(j)}:=\Tr_{S_1\cdots S_{j-1}}\Gamma_j
                 =\Gamma_n^{S_j}
\]
its marginal on $S_j$.

Let $K$ be an $n$-dimensional classical register with orthonormal basis
$\{|k\rangle\}_{k=1}^n$, and define the catalyst on
$C=S^{\otimes(n-1)}K$ by
\begin{equation}
    \tau^C
    :=\frac1n\sum_{k=1}^n
    \rho^{\otimes(k-1)}\otimes\Gamma_{n-k}
    \otimes|k\rangle\!\langle k|^K.
\label{eq:catalyst}
\end{equation}
Thus
\begin{equation}
    \rho\otimes\tau^C
    =\frac1n\sum_{k=1}^n
    \rho^{\otimes k}\otimes\Gamma_{n-k}
    \otimes|k\rangle\!\langle k|^K.
\label{eq:input}
\end{equation}

Conditioned on $K$, apply the identity for $k<n$ and $\Lambda$ for
$k=n$. By Assumption \textup{(A1)}, this is a free operation. Applied
to~\eqref{eq:input}, it gives
\begin{equation}
    \frac1n\sum_{k=1}^{n-1}
    \rho^{\otimes k}\otimes\Gamma_{n-k}
    \otimes|k\rangle\!\langle k|
    +\frac1n\Gamma_n\otimes|n\rangle\!\langle n|.
\label{eq:conditioned}
\end{equation}

Next apply the cyclic permutation
\[
    S_1S_2\cdots S_n\longmapsto S_nS_1\cdots S_{n-1}
\]
and cyclically relabel the classical register,
\[
    |k\rangle\longmapsto |k+1\rangle\quad(k<n),
    \qquad
    |n\rangle\longmapsto|1\rangle.
\]
Both operations are free by Assumptions \textup{(A3)} and \textup{(A2)}. Let
$\omega^{SC}$ denote the resulting state, where the first $S$ is the system output and the remaining $n-1$ $S$ together with $K$ form the catalyst.

Tracing out the system output yields
\begin{align}
    \omega^C
    &=\frac1n\sum_{k=1}^{n-1}
      \rho^{\otimes k}\otimes\Gamma_{n-k-1}
      \otimes|k+1\rangle\!\langle k+1|
      +\frac1n\Gamma_{n-1}\otimes|1\rangle\!\langle1| \\
    &=\frac1n\sum_{\ell=1}^{n}
      \rho^{\otimes(\ell-1)}\otimes\Gamma_{n-\ell}
      \otimes|\ell\rangle\!\langle\ell|
      =\tau^C.
\label{eq:return}
\end{align}
Hence the catalyst is returned exactly.

The system marginal is
\begin{equation}
    \omega^S
    =\frac1n\sum_{k=1}^{n-1}\Gamma_{n-k}^{(n-k)}
     +\frac1n\Gamma_n^{(n)}
    =\frac1n\sum_{j=1}^n\Gamma_j^{(j)}.
\label{eq:system}
\end{equation}
Therefore, using~\eqref{eq:good} and the bound
$\|\eta-\sigma\|_1\le2$ for any states $\eta$ and $\sigma$,
\begin{align}
    \|\omega^S-\sigma\|_1
    &\le \frac1n\sum_{j=1}^n
        \|\Gamma_j^{(j)}-\sigma\|_1 \\
    &< \frac{m}{n}\,\varepsilon
       +2\frac{n-m}{n} \\
    &< \varepsilon+2\delta.
\label{eq:error0}
\end{align}
Since $\varepsilon,\delta>0$ are arbitrary, the system output can be
made arbitrarily close to $\sigma$ while the catalyst is returned
exactly. Hence
\[
    \rho\xrightarrow[\mathcal O]{\cc}\sigma.
\]
\end{proof}

\section{Learning and Production Protocol}

\begin{assumption}[Full-rank states and informational completeness]
\label{ass:prop2}
For the converse direction we additionally use the following
assumptions.

\begin{enumerate}[label=\textup{(A\arabic*)},leftmargin=3.2em]

\item[(A4)] \emph{Full-rank free states.}
Every finite-dimensional system $C$ admits a full-rank free state,
\[
    \exists\,\omega^C\in\mathcal F(C)
    \qquad\text{such that}\qquad
    \omega^C>0.
\]

\item[(A5)] \emph{Free informationally complete (IC) measurements.}
Every finite-dimensional system $B$ admits a finite
informationally complete POVM
$\{M_z^B\}_{z\in\mathcal Z}$ whose measurement channel
\[
    \mathcal M_B(X)
    :=
    \sum_{z\in\mathcal Z}
    \operatorname{Tr}(M_z^B X)
    |z\rangle\!\langle z|
\]
belongs to
\[
    \mathcal O(B\to Z),
\]
and
\[
    \operatorname{span}_{\mathbb R}
    \{M_z^B:z\in\mathcal Z\}
    =
    \operatorname{Herm}(B).
\]

\end{enumerate}
\end{assumption}

\begin{lemma}~\cite[Sec.~5.1.4]{BeigiThesis}
\label{lem:ic}
If $\{M_z\}_{z\in\mathcal Z}$ is an IC POVM on a finite-dimensional
system $B$, then there exist Hermitian operators $\{F_z\}_{z\in\mathcal Z}$
such that, for every $X\in\operatorname{Herm}(B)$,
\begin{equation}
    X=\sum_{z\in\mathcal Z}\operatorname{Tr}(M_zX)F_z .
    \label{eq:ic-reconstruction}
\end{equation}
Hence, with $K_M:=\max_z\norm{F_z}$,
\begin{equation}
    \norm{X}
    \le
    K_M\sum_{z\in\mathcal Z}
    \left|\operatorname{Tr}(M_zX)\right|.
    \label{eq:ic-bound}
\end{equation}
\end{lemma}

\begin{lemma}[Blackwell--Dubins~\cite{BlackwellDubins1962}]
\label{lem:bd}
Let $P,Q$ be probability distributions on $\mathcal Z^{\mathbb N}$,
where $\mathcal Z$ is finite and $Q\ll P$
(i.e., $P(E)=0$ implies $Q(E)=0$).
Then
\begin{equation}
\sum_{\mathrm{history}\in\mathcal Z^k}
Q(\mathrm{history})
\sup_{s\ge1}
\sum_{\mathrm{future}\in\mathcal Z^s}
\left|
P(\mathrm{future}\mid\mathrm{history})
-Q(\mathrm{future}\mid\mathrm{history})
\right|
\xrightarrow[k\to\infty]{}0.
\label{eq:bd}
\end{equation}
When the observed history is sampled according to \(Q\), the predictions of \(P\) and \(Q\) for the entire future become arbitrarily close on average as the history grows. 
\end{lemma}
\begin{proposition}
In a QRT with tensor-product closure and assumptions
\textup{(A1)}, \textup{(A2)}, \textup{(A4)}, and \textup{(A5)},
\[
 \rho\freeccto{\cO}\sigma
 \quad\Longrightarrow\quad R_m^{\cO}(\rho\to\sigma)\ge1.
\]
\label{prop:converse}
\end{proposition}

\begin{proof}
Fix $\varepsilon>0$ and a correlated-catalytic protocol
$\Lambda\in\cO(AC\to BC)$ with catalyst $\tau^C$ such that
\begin{equation}
 \mu^{BC}=\Lambda(\rho^A\otimes\tau^C),\qquad
 \mu^C=\tau^C,\qquad \norm{\mu^B-\sigma^B}<\varepsilon.
 \label{eq:cc}
\end{equation}
Choose a full-rank free state $\omega^C$ such that $\omega^C\ge c\tau^C$, with $0<c<1$,
and a free informationally complete POVM $\{M_z\}_{z\in\mathcal Z}$
on $B$, with a constant $K_M$ satisfying
\[
 \norm{X}\le K_M\sum_z|\Tr(M_zX)|\qquad(X=X^\dagger).
\]

Each round feeds a fresh copy of $\rho$ and the state of the previous round in the memory $C$ into
$\Lambda$, the output register $B$ is measured using the POVM $\{M_z\}_{z\in\mathcal Z}$, and $C$ is passed to the next round.
The circuit acting on $A^{\otimes k}C$ is free by tensor-product
closure, composition, and \textup{(A5)}.

\begin{center}
\begin{tikzpicture}[x=1cm,y=1cm,line width=.65pt,
  every node/.style={font=\small},line cap=round,line join=round]
  \foreach \x/\j in {1.8/1,5.3/2,9.5/k} {
    \draw (\x,-.25) rectangle ++(.72,1.10);
    \node at (\x+.36,.30) {$\Lambda$};
    \draw (\x-.70,.60)--(\x,.60);
    \node[above,inner sep=2pt] at (\x-.36,.60) {$\rho^A$};
    \draw (\x+.72,.60)--(\x+1.02,.60);
    \draw (\x+1.02,.34) rectangle ++(.54,.52);
    \draw (\x+1.11,.47) arc[start angle=180,end angle=0,radius=.18];
    \draw (\x+1.29,.46)--(\x+1.42,.69);
    \draw[double,double distance=.6pt] (\x+1.56,.60)--(\x+2.12,.60);
    \node[above,inner sep=2pt] at (\x+1.85,.60) {$z_{\j}$};
  }
  \node[left,inner sep=3pt] at (.65,0) {$\omega\ \text{or}\ \tau$};
  \draw (.65,0)--(1.8,0);
  \draw (2.52,0)--(5.3,0);
  \draw (6.02,0)--(9.5,0);
  \draw (10.22,0)--(11.8,0);
  \node[right,inner sep=3pt] at (11.8,0) {$C$};
  \node at (8.30,.60) {$\cdots$};
  \draw[decorate,decoration={brace,mirror,amplitude=4pt}]
     (1.10,-.43)--(11.62,-.43);
  \node at (6.36,-.84) {$k\text{ rounds}$};
  \node at (6.36,-1.31) {$\Updownarrow$};
  \draw (5.25,-2.65) rectangle (7.47,-1.72);
  \node at (6.36,-2.185) {$k\text{ rounds}$};
  \draw (3.95,-1.93)--(5.25,-1.93);
  \node[left,inner sep=4pt] at (3.95,-1.93) {$\rho^{\otimes k}$};
  \draw (3.95,-2.43)--(5.25,-2.43);
  \node[left,inner sep=4pt] at (3.95,-2.43) {$\omega\ \text{or}\ \tau$};
  \draw[double,double distance=.6pt] (7.47,-1.93)--(8.78,-1.93);
  \node[right,inner sep=4pt] at (8.78,-1.93) {$h=(z_1,\ldots,z_k)$};
  \draw (7.47,-2.43)--(8.78,-2.43);
  \node[right,inner sep=4pt] at (8.78,-2.43) {$\omega_h\ \text{or}\ \tau_h$};
\end{tikzpicture}
\end{center}

Initial memories $\omega$ and $\tau$ give the ensembles
$\{P(h),\omega_h\}_h$ and $\{Q(h),\tau_h\}_h$, respectively. We have
\begin{equation}
 P(h)\ge cQ(h),\qquad \sum_hQ(h)\tau_h=\tau.
 \label{eq:histories}
\end{equation}
The same domination holds at every length, so the laws of the
infinite measurement sequences satisfy $P\ge cQ$ and hence $Q\ll P$.

Run the circuit starting with the free state $\omega$ in $C$. Depending on the observed $h$, toss a
coin with success probability $a(h)=cQ(h)/P(h)$, setting $a(h)=0$
when $P(h)=0$. Keep $\omega_h$ on success and reset the memory
to $\omega$ on failure. Thus
\begin{equation}
 \{P(h),\omega_h\}_h
 \longmapsto
 \bigl\{(cQ(h),\omega_h)\bigr\}_h\cup\bigl\{(1-c,\omega)\bigr\}.
 \label{eq:filter}
\end{equation}
Conditioning on success changes the history distribution from $P$ to
$Q$, while leaving the state $\omega_h$ unchanged.

Repeat $T$ times in parallel, keeping the memory from the first successful
attempt. If all attempts fail, reset it to $\omega$. The probability
that the first success occurs at attempt $j$ with history $h$ is
$(1-c)^{j-1}cQ(h)$. Then, the resulting ensemble is
\begin{equation}
 \bigl\{((1-f_T)Q(h),\omega_h)\bigr\}_h
 \ \cup\ \bigl\{(f_T,\omega)\bigr\},
 \label{eq:branches}
\end{equation}
with $f_T=(1-c)^T$.

For each branch of this ensemble, run $m$ further rounds of $\Lambda$, retain $B_1,\ldots,B_m$, and discard the final memory:
\begin{center}
\begin{tikzpicture}[x=1cm,y=1cm,line width=.65pt,
  every node/.style={font=\small},line cap=round,line join=round]
  \foreach \x/\j in {1.8/1,5.3/2,9.5/m} {
    \draw (\x,-.25) rectangle ++(.72,1.10);
    \node at (\x+.36,.30) {$\Lambda$};
    \draw (\x-.70,.60)--(\x,.60);
    \node[above,inner sep=2pt] at (\x-.36,.60) {$\rho^A$};
    \draw (\x+.72,.60)--(\x+1.12,.60)--(\x+1.12,1.35);
    \node[above,inner sep=3pt] at (\x+1.12,1.35) {$B_{\j}$};
  }
  \node[left,inner sep=3pt] at (.65,0) {$\omega_h\ \text{or}\ \omega$};
  \draw (.65,0)--(1.8,0);
  \draw (2.52,0)--(5.3,0);
  \draw (6.02,0)--(9.5,0);
  \draw (10.22,0)--(11.65,0);
  \draw (11.65,-.20)--(11.65,.20);
  \draw (11.76,-.14)--(11.76,.14);
  \draw (11.87,-.08)--(11.87,.08);
  \node[right,inner sep=4pt] at (11.95,0) {$\Tr_C$};
  \node at (8.30,.60) {$\cdots$};
  \draw[decorate,decoration={brace,mirror,amplitude=4pt}]
     (1.10,-.43)--(11.87,-.43);
  \node at (6.48,-.84) {$m\text{ rounds}$};
\end{tikzpicture}
\end{center}
Write $\Gamma_{\omega_h}$ and $\Gamma_\omega$ for the outputs starting from
$\omega_h$ and $\omega$, respectively. Summing over success and failure
branches gives, for every $1\le i\le m$,
\begin{equation}
 \Gamma^{B_i}=(1-f_T)\sum_hQ(h)\Gamma_{\omega_h}^{B_i}
                  +f_T\Gamma_\omega^{B_i}.
 \label{eq:output}
\end{equation}
Hypothetically measuring the future outputs and summing over the
other outcomes gives
\begin{equation}
\Tr(M_z\Gamma_{\omega_h}^{B_i})=P(Z_{k+i}=z\mid h).
 \label{eq:actual-statistics}
\end{equation}
Starting from $\tau$, exact catalyst return implies that each
unconditional output marginal is $\mu^B$. Hence
\begin{equation}
 \sum_hQ(h)Q(Z_{k+i}=z\mid h)=\Tr(M_z\mu^B).
 \label{eq:reference-statistics}
\end{equation}
Using informational completeness and including the failure branch,
\begin{align}
 \norm{\Gamma^{B_i}-\sigma^B}
 &\le \varepsilon+\norm{\Gamma^{B_i}-\mu^B}\notag\\
 &\le \varepsilon+(1-f_T)
       \left\lVert\sum_hQ(h)\Gamma_{\omega_h}^{B_i}-\mu^B\right\rVert_1+2f_T\notag\\
 &\le \varepsilon+(1-f_T)K_M\sum_z
       \left|\sum_hQ(h)\bigl[P(Z_{k+i}=z\mid h)
                         -Q(Z_{k+i}=z\mid h)\bigr]\right|+2f_T\notag\\
 &\le \varepsilon+K_M\sum_hQ(h)\sum_z
       |P(Z_{k+i}=z\mid h)-Q(Z_{k+i}=z\mid h)|+2f_T.
 \label{eq:error}
\end{align}

Since $Q\ll P$, the Blackwell--Dubins theorem~\cite{BlackwellDubins1962}
and bounded convergence imply
\begin{equation}
 \eta_k:=\sum_{h\in\mathcal Z^k}Q(h)\sup_{\ell\ge1}
       \sum_{u\in\mathcal Z^\ell}|P(u\mid h)-Q(u\mid h)|
       \xrightarrow[k\to\infty]{}0,
 \label{eq:bd0}
\end{equation}
where $u$ denotes the next $\ell$ measurement outcomes after $h$.
Marginalising these future distributions therefore yields
\begin{equation}
 \max_{1\le i\le m}\norm{\Gamma^{B_i}-\sigma^B}
 \le\varepsilon+K_M\eta_k+2(1-c)^T.
 \label{eq:uniform}
\end{equation}
Choose $k,T$ so that $K_M\eta_k<\varepsilon$ and
$2(1-c)^T<\varepsilon$. Both choices are independent of $m$.
The entire protocol is free by \textup{(A1)}, \textup{(A2)},
\textup{(A4)}, \textup{(A5)}, tensor-product closure, and composition.
It consumes at most $Tk+m$ copies of $\rho$; unused inputs are discarded.
Consequently,
\[
 \max_{1\le i\le m}\norm{\Gamma^{B_i}-\sigma^B}<3\varepsilon,
 \qquad \frac{m}{Tk+m}\xrightarrow[m\to\infty]{}1.
\]
As $\varepsilon>0$ was arbitrary, $R_m^{\cO}(\rho\to\sigma)\ge1$.
\end{proof}

Together, Propositions~\ref{prop:cyclic} and~\ref{prop:converse} give the
main result.

\begin{theorem}[Catalytic--marginal equivalence]
\label{thm:equivalence}
In a QRT with tensor-product closure and assumptions \textup{(A1)},
\textup{(A2)}, \textup{(A3)}, \textup{(A4)}, and \textup{(A5)},
\begin{equation}
    \rho\freeccto{\cO}\sigma
    \quad\Longleftrightarrow\quad
    \Rm^{\cO}(\rho\to\sigma)\ge1 .
    \label{eq:equivalence}
\end{equation}
\end{theorem}
\begin{proof}
This follows from Propositions~\ref{prop:cyclic} and~\ref{prop:converse}.
\end{proof}

\section{Rate consequences and applications}

This section derives consequences of Theorem~\ref{thm:equivalence} for
asymptotic rates and applies them to concrete resource theories.

\subsection*{Supermultiplicativity of $\Rm^{\cO}$ and correlated catalysis transitivity}

\begin{lemma}[Supermultiplicativity]
\label{lem:composition}
In a QRT with tensor-product closure and \textup{(A3)}, if
$\Rm^{\cO}(\rho\to\sigma)$ and $\Rm^{\cO}(\sigma\to\omega)$ are positive,
then
\begin{equation}
    \Rm^{\cO}(\rho\to\omega)
    \ge\Rm^{\cO}(\rho\to\sigma)\,\Rm^{\cO}(\sigma\to\omega).
    \label{eq:composition}
\end{equation}
\end{lemma}
\begin{proof}
Fix $\varepsilon>0$, 
\[
0<r_1<\Rm(\rho\to\sigma),\quad \text{and} \quad
0<r_2<\Rm(\sigma\to\omega).
\]
Choose a free protocol $\Psi$ taking $k$ copies of $\sigma$ to $\ell$
outputs such that
\[
\frac{\ell}{k}>r_2,
\qquad
\norm{\Psi(\sigma^{\otimes k})^{C_j}-\omega}<\varepsilon
\quad
\forall j.
\]
For this fixed $k$, choose a free protocol
\[
\Gamma:=\Lambda(\rho^{\otimes n})^{B_1\cdots B_m}
\]
with
\[
\frac{m}{n}>r_1,
\qquad
\norm{\Gamma^{B_t}-\sigma}<\frac{\varepsilon}{k}
\quad
\forall t.
\]

The outputs $B_1,\ldots,B_m$ may be correlated, so they cannot in general
be fed directly into $\Psi$. Instead, run $\Lambda$ independently $k$
times and group together the $t$-th outputs. The $t$-th group has state
$(\Gamma^{B_t})^{\otimes k}$, and
\[
\norm{(\Gamma^{B_t})^{\otimes k}-\sigma^{\otimes k}}
\le
k\norm{\Gamma^{B_t}-\sigma}
<\varepsilon.
\]
Applying $\Psi$ to each group, contractivity and the triangle inequality
show that every final marginal is $2\varepsilon$-close to $\omega$.

The resulting free protocol uses $kn$ copies of $\rho$ and produces
$m\ell$ outputs, hence has rate
\[
\frac{m\ell}{kn}
=
\frac{m}{n}\frac{\ell}{k}
>
r_1r_2.
\]
Since $\varepsilon$ is arbitrary and $r_1,r_2$ can approach the two
marginal rates,
\[
\Rm(\rho\to\omega)
\ge
\Rm(\rho\to\sigma)\Rm(\sigma\to\omega).
\]
\end{proof}

Together with Theorem~\ref{thm:equivalence}, this makes correlated catalysis transitive:
\begin{equation}
    \rho\freeccto{\cO}\sigma,\quad\sigma\freeccto{\cO}\omega
    \quad\Longrightarrow\quad\rho\freeccto{\cO}\omega .
    \label{eq:transitivity-sm}
\end{equation}

\subsection*{Equality of asymptotic rates}
\begin{definition}[Other asymptotic rates]
\label{def:rates}
The \emph{catalytic rate} $\Rc^{\cO}(\rho\to\sigma)$ is defined as in
Definition~\ref{def:Rm}, except that the free map
$\Lambda\in\cO(A^{\otimes n}C\to B^{\otimes m}C)$ also acts on a
finite-dimensional catalyst state $\tau^C$ (which may depend on
$\varepsilon,\delta$), and the output
$\Omega:=\Lambda(\rho^{\otimes n}\otimes\tau^C)$ must satisfy
\[
    \Omega^C=\tau^C,
    \qquad
    \norm{\Omega^{B_1\cdots B_m}-\sigma^{\otimes m}}<\varepsilon .
\]
The \emph{marginal catalytic rate} $\Rmc^{\cO}(\rho\to\sigma)$ is
defined similarly, but with the conditions
\[
    \Omega^C=\tau^C,
    \qquad
    \max_{1\le i\le m}
    \norm{\Omega^{B_i}-\sigma}<\varepsilon .
\]
The \emph{asymptotic rate} $R^{\cO}(\rho\to\sigma)$ is defined as in
Definition~\ref{def:Rm}, with a free map
$\Lambda\in\cO(A^{\otimes n}\to B^{\otimes m})$
whose output $\Omega:=\Lambda(\rho^{\otimes n})$ satisfies
\[
    \norm{\Omega-\sigma^{\otimes m}}<\varepsilon .
\]
\end{definition}

\begin{corollary}[Equality of rates]
\label{cor:rates-sm}
Under the hypotheses of Theorem~\ref{thm:equivalence},
\begin{equation}
    \Rm^{\cO}(\rho\to\sigma)=\Rc^{\cO}(\rho\to\sigma)=\Rmc^{\cO}(\rho\to\sigma).
    \label{eq:rates-sm}
\end{equation}
\end{corollary}
\begin{proof}
To prove this corollary, we will show the following set of inequalities: $\Rc\le\Rmc\le\Rm\le\Rc$.

\emph{Step 1: $\Rc\le\Rmc$.}
If the whole output is $\varepsilon$-close to $\sigma^{\otimes m}$, so is
every single output, since
$\norm{\Omega^{B_i}-\sigma}\le\norm{\Omega^{B_1\cdots B_m}-\sigma^{\otimes m}}$.

\emph{Step 2: $\Rmc\le\Rm$.}
Take a protocol admissible for $\Rmc^{\cO}$ with marginal errors below
$\varepsilon$, and call its output $\Gamma:=\Omega^{B_1\cdots B_m}$.
Because the catalyst is returned exactly, this protocol is itself a
correlated-catalytic conversion $\rho^{\otimes n}\ccto{\cO}\Gamma$.
Proposition~\ref{prop:converse} gives
$\Rm^{\cO}(\rho^{\otimes n}\to\Gamma)\ge1$: for any $\delta>0$, a free map
without catalyst turns $N$ copies of $\rho^{\otimes n}$ into
$L>(1-\delta)N$ blocks, each $\varepsilon$-close to $\Gamma$. Splitting every
block into its $m$ outputs gives $Lm$ outputs, each $2\varepsilon$-close to
$\sigma$, from $Nn$ copies of $\rho$. So the rate
$Lm/(Nn)>(1-\delta)\,m/n$ is reached without a catalyst.

\emph{Step 3: $\Rm\le\Rc$.}
Let $a,b\in\mathbb N$ with $b/a<\Rm^{\cO}(\rho\to\sigma)$. Reading one copy
of $\rho^{\otimes a}$ as $a$ copies of $\rho$, and $b$ copies of $\sigma$ as
one copy of $\sigma^{\otimes b}$, gives
\[
    \Rm^{\cO}(\rho^{\otimes a}\to\rho)\ge a,
    \qquad
    \Rm^{\cO}(\sigma\to\sigma^{\otimes b})\ge\frac1b .
\]
Applying Lemma~\ref{lem:composition} twice,
\[
    \Rm^{\cO}(\rho^{\otimes a}\to\sigma^{\otimes b})
    \ge a\cdot\Rm^{\cO}(\rho\to\sigma)\cdot\frac1b>1,
\]
so Proposition~\ref{prop:cyclic} gives
$\rho^{\otimes a}\ccto{\cO}\sigma^{\otimes b}$. That is, for every
$\varepsilon>0$ a catalytic protocol takes $a$ copies of $\rho$ to $b$
outputs that are jointly $\varepsilon$-close to $\sigma^{\otimes b}$, which
is what Definition~\ref{def:rates} requires for $\Rc^{\cO}$ with $n=a$ and
$m=b$. Hence $\Rc\ge b/a$, and letting $b/a\uparrow\Rm$ gives
$\Rc\ge\Rm$.
\end{proof}

\subsection*{Reference states}

By Lemma~\ref{lem:composition}, every state $\xi$ satisfies
\[
    \Rm^{\cO}(\xi\to\xi)\ge1,
    \quad \text{and} \quad
    \Rm^{\cO}(\xi\to\xi)\ge\Rm^{\cO}(\xi\to\xi)^2.
\]
Thus $\Rm^{\cO}(\xi\to\xi)\in\{1,\infty\}$. A \emph{reference state} is a state $\varphi$ with $\Rm^{\cO}(\varphi\to\varphi)=1$. Fix such a state and set
\begin{equation}
    \Dm(\rho):=\Rm^{\cO}(\rho\to\varphi),
    \qquad
    \Cm(\rho):=\frac{1}{\Rm^{\cO}(\varphi\to\rho)}.
    \label{eq:reference}
\end{equation}
Lemma~\ref{lem:composition} through
$\rho$ gives $\Dm(\rho)\le\Cm(\rho)$; this is trivial if $\Dm(\rho)=0$ or
$\Cm(\rho)=\infty$.

\begin{theorem}[Reference-state rate formula]
\label{thm:reference}
Under the hypotheses of Theorem~\ref{thm:equivalence}, let
$0<\Dm(\rho)<\infty$ and $0<\Cm(\sigma)<\infty$, and suppose that
$\Dm(\rho)=\Cm(\rho)$ or $\Dm(\sigma)=\Cm(\sigma)$. Then
\begin{equation}
    \Rm^{\cO}(\rho\to\sigma)=\Rc^{\cO}(\rho\to\sigma)
    =\Rmc^{\cO}(\rho\to\sigma)=\frac{\Dm(\rho)}{\Cm(\sigma)} .
    \label{eq:reference-rate}
\end{equation}
\end{theorem}
\begin{proof}
Composition through $\varphi$ gives
$\Rm(\rho\to\sigma)\ge\Dm(\rho)/\Cm(\sigma)$. Composition through $\sigma$
and through $\rho$ gives
\[
    \Dm(\rho)\ge\Rm(\rho\to\sigma)\,\Dm(\sigma),
    \qquad
    \frac{1}{\Cm(\sigma)}\ge\frac{\Rm(\rho\to\sigma)}{\Cm(\rho)} .
\]
Substituting $\Dm(\sigma)=\Cm(\sigma)$ into the first inequality, or
$\Cm(\rho)=\Dm(\rho)$ into the second, gives the reverse bound.
Corollary~\ref{cor:rates-sm} gives the remaining equalities.
\end{proof}

\subsection*{Bipartite LOCC}

LOCC satisfies the hypotheses of
Theorem~\ref{thm:equivalence}. Choose the Bell pair $\Phi_2$ as the reference
state, let $\Ed$ be the distillable entanglement in ebits, and let
$S(\phi_A)=-\Tr(\phi_A\log_2\phi_A)$ for a bipartite pure state $\phi$.

For every bipartite state $\rho$ and every entangled pure state $\phi$, Ref.~\cite{ganardi2024} establishes
\begin{equation}
    \Rm^{\mathrm{LOCC}}(\rho\to\phi)=R^{\mathrm{LOCC}}(\rho\to\phi)
    =\frac{\Ed(\rho)}{S(\phi_A)} .
    \label{eq:locc-known}
\end{equation}
In particular, $\Rm^{\mathrm{LOCC}}(\Phi_2\to\Phi_2)=1$,
$\Dm(\rho)=\Ed(\rho)$, and $\Dm(\phi)=\Cm(\phi)=S(\phi_A)$.
Corollary~\ref{cor:rates-sm} therefore yields
\begin{equation}
    \Rm^{\mathrm{LOCC}}(\rho\to\phi)
    =\Rc^{\mathrm{LOCC}}(\rho\to\phi)
    =\Rmc^{\mathrm{LOCC}}(\rho\to\phi)
    =R^{\mathrm{LOCC}}(\rho\to\phi)
    =\frac{\Ed(\rho)}{S(\phi_A)},
    \label{eq:locc-pure-target}
\end{equation}
and Theorem~\ref{thm:equivalence} gives the one-shot criterion
\begin{equation}
    \boxed{\rho\freeccto{\mathrm{LOCC}}\phi
    \quad\Longleftrightarrow\quad \Ed(\rho)\ge S(\phi_A).}
    \label{eq:locc-one-shot}
\end{equation}
Taking $\phi=\Phi_2$ shows that correlated catalysts do not increase the
Bell-pair distillation rate. More generally, if
$\Rm^{\mathrm{LOCC}}(\rho\to\sigma)$ and $\Ed(\sigma)$ are positive,
Lemma~\ref{lem:composition} gives
\begin{equation}
    \Ed(\rho)=\Rm^{\mathrm{LOCC}}(\rho\to\Phi_2)
    \ge\Rm^{\mathrm{LOCC}}(\rho\to\sigma)\,\Ed(\sigma).
    \label{eq:ed-composition}
\end{equation}
Since $\rho\ccto{\mathrm{LOCC}}\sigma$ means
$\Rm^{\mathrm{LOCC}}(\rho\to\sigma)\ge1$, it implies
$\Ed(\rho)\ge\Ed(\sigma)$ (trivially if $\Ed(\sigma)=0$). In
particular, a state with zero distillable entanglement cannot be converted by
correlated catalysis into any fixed entangled pure state.

For an entangled pure source $\psi$ and a target $\sigma$ with
$0<\Cm(\sigma)<\infty$, Theorem~\ref{thm:reference} instead gives
\begin{equation}
    \Rm^{\mathrm{LOCC}}(\psi\to\sigma)
    =\Rc^{\mathrm{LOCC}}(\psi\to\sigma)
    =\Rmc^{\mathrm{LOCC}}(\psi\to\sigma)
    =\frac{S(\psi_A)}{\Cm(\sigma)} .
    \label{eq:locc-pure-source}
\end{equation}
Here $\Cm$ is the marginal cost~\eqref{eq:reference}; it is not identified
with the ordinary entanglement cost. 

The marginal cost is positive for every entangled state. Indeed, squashed entanglement $E_{\mathrm{sq}}$ is LOCC-monotone, strongly superadditive, continuous, and faithful~\cite{christandl2004,alicki2004,brandao2011}. Hence, for any marginal conversion of $n$ Bell pairs into $m$ approximate copies of $\sigma$,
\[
    n\ge\sum_{i=1}^m E_{\mathrm{sq}}(\Omega^{A_iB_i})
    \ge m\bigl(E_{\mathrm{sq}}(\sigma)-o(1)\bigr).
\]
Consequently, every entangled state $\sigma$ satisfies
\begin{equation}
    0<E_{\mathrm{sq}}(\sigma)
    \le\Cm(\sigma)\le E_C(\sigma)<\infty,
\end{equation}
where $E_C$ is the ordinary entanglement cost~\cite{hayden2001}.

\section{Secret key and private communication}

\begin{definition}[Private states]
\label{def:private-state}
Let $D=K_AK_BS_AS_B$, with Alice holding $K_AS_A$ and Bob holding
$K_BS_B$, and let $\dim K_A=\dim K_B=d$. A private state with key
dimension $d$ is a state of the form~\cite{horodecki2009private}
\begin{equation}
    \gamma_d
    =\frac{1}{d}\sum_{i,j=0}^{d-1}
      |ii\rangle\!\langle jj|^{K_AK_B}
      \otimes U_i\theta^{S_AS_B}U_j^\dagger,
    \label{eq:private-state-definition}
\end{equation}
where $\theta^{S_AS_B}$ is a density operator and each $U_i$ is a
unitary on $S_AS_B$. The systems $K_A,K_B$ are the key registers and
$S_A,S_B$ are the shield registers.
For every extension of $\gamma_d$ to an external system $E$, measuring
the key registers in the displayed bases and discarding the shield gives
an ideal secret key,
\begin{equation}
    \kappa_d^{K_AK_BE}
    =\frac{1}{d}\sum_{i=0}^{d-1}
      |ii\rangle\!\langle ii|^{K_AK_B}\otimes\eta^E,
    \label{eq:ideal-secret-key-definition}
\end{equation}
for a density operator $\eta^E$. Thus the outcomes agree, are uniform,
and are independent of $E$, which may include Eve's quantum side
information and the public communication record.
\end{definition}

\begin{definition}[Distillable secret key]
\label{def:distillable-secret-key}
For a bipartite state $\rho^{AB}$, define
\begin{equation}
    K_D(\rho)
    :=\sup_{\{\Lambda_n,\gamma_{d_n}\}}
      \liminf_{n\to\infty}\frac{\log_2d_n}{n},
    \label{eq:distillable-secret-key-definition}
\end{equation}
where the supremum runs over sequences of LOCC maps $\Lambda_n$
acting on $\rho^{\otimes n}$ and private states $\gamma_{d_n}$ such that
\begin{equation}
    \lim_{n\to\infty}
    \norm{\Lambda_n(\rho^{\otimes n})-\gamma_{d_n}}=0.
\end{equation}
The output bipartition is $K_AS_A:K_BS_B$.
This equals the optimal rate of classical secret-key distillation
when Eve holds a purification of the input and receives the public
communication record~\cite{horodecki2009private}.
\end{definition}

\begin{definition}[Two-way private-state generation capacity]
\label{def:two-way-private-capacity}
An $n$-use protocol $\Pi_n$ for a channel $\mathcal N:A'\to B'$ starts
with locally prepared product ancillas and interleaves $n$ uses of
$\mathcal N$ with LOCC, allowing unlimited two-way public classical
communication. Let $\omega_n^{D_n}$ be its unconditional output.
Define
\begin{equation}
    P_2(\mathcal N)
    :=\sup_{\{\Pi_n,\gamma_{d_n}\}}
      \liminf_{n\to\infty}\frac{\log_2d_n}{n},
    \label{eq:two-way-private-capacity-definition}
\end{equation}
where the supremum is restricted to sequences satisfying
\begin{equation}
    \lim_{n\to\infty}\norm{\omega_n^{D_n}-\gamma_{d_n}}=0.
\end{equation}
This capacity equals the two-way secret-key capacity of
$\mathcal N$~\cite{horodecki2009private,pirandola2017}.
\end{definition}

\begin{definition}[Correlated-catalytic private-state generation capacity]
\label{def:cc-private-capacity}
For each $n$, let $\tau_n^{C_n}$ be a bipartite catalyst on
$C_n=C_{A,n}C_{B,n}$. An $n$-use adaptive protocol implements a map
$\mathcal T_n:C_n\to D_nC_n$ using $n$ calls to $\mathcal N$
interleaved with LOCC. All ancillary registers are locally prepared
and initially uncorrelated with the catalyst. Write
\begin{equation}
    \Omega_n^{D_nC_n}:=\mathcal T_n(\tau_n^{C_n}).
\end{equation}
Define
\begin{equation}
    P_{2,\mathrm{cc}}(\mathcal N)
    :=\sup_{\{\tau_n,\mathcal T_n,\gamma_{d_n}\}}
      \liminf_{n\to\infty}\frac{\log_2d_n}{n},
    \label{eq:cc-private-capacity-definition}
\end{equation}
where the supremum is restricted by
\begin{equation}
    \Omega_n^{C_n}=\tau_n^{C_n}\quad\text{for every }n,
    \qquad
    \lim_{n\to\infty}\norm{\Omega_n^{D_n}-\gamma_{d_n}}=0.
\end{equation}
The entire bipartite catalyst is returned exactly, while correlations
between $D_n$ and $C_n$ are allowed.
\end{definition}

In these definitions, all output, shield, and catalyst systems are
finite-dimensional for each $n$, but their dimensions may grow with
$n$ without a uniform bound. The rates are measured in secret bits
per input copy for $K_D$ and per channel use for $P_2$ and
$P_{2,\mathrm{cc}}$.

\emph{States.}
By Theorem~\ref{thm:equivalence}, $\rho\ccto{\mathrm{LOCC}}\sigma$ gives free protocols $\Gamma=\Lambda(\rho^{\otimes n})$ with vanishing marginal error and $m/n\to1$. Weak additivity and LOCC monotonicity of $K_D$ follow from its operational definition~\cite{horodecki2009private}. Together with strong superadditivity~\cite{wolf2006} and lower semicontinuity, they give
\begin{equation}
    nK_D(\rho)=K_D(\rho^{\otimes n})
    \ge K_D(\Gamma)
    \ge\sum_{i=1}^mK_D(\Gamma^{D_i})
    \ge m\bigl[K_D(\sigma)-o(1)\bigr].
    \label{eq:key-chain}
\end{equation}
Dividing by $n$ and taking the limits yields $K_D(\rho)\ge K_D(\sigma)$.

\emph{Channels.} Write $P_2(\mathcal N)$ for the two-way private-state generation capacity and $P_{2,\mathrm{cc}}(\mathcal N)$ for its correlated-catalytic counterpart~\cite{horodecki2009private,pirandola2017}. Fix an adaptive catalytic protocol using $n$ calls to $\mathcal N$. Its entire circuit is a CPTP map $\mathcal T:C\to DC$ of the form
\begin{equation}
    \mathcal T
    =\mathcal L_n\circ(\mathcal N\otimes\id)
     \circ\mathcal L_{n-1}\circ\cdots
     \circ(\mathcal N\otimes\id)\circ\mathcal L_0,
    \label{eq:adaptive-channel-map}
\end{equation}
where each $\mathcal L_j$ is LOCC and may depend on earlier measurement outcomes, and the identities act on retained memories. 

For the catalyst $\tau^C$, write $\Omega^{DC}:=\mathcal T(\tau^C)$. The catalytic task requires 
\[     \Omega^C=\tau^C,     \qquad     \norm{\Omega^D-\gamma_d}\le\varepsilon, \] where $\gamma_d$ is a private state with key dimension $d$. The output $D=K_AK_BS_AS_B$ contains the key registers $K_A,K_B$ and shield registers $S_A,S_B$. The capacity $P_{2,\mathrm{cc}}(\mathcal N)$ is the supremum of asymptotically achievable rates $(\log_2d)/n$ as $n\to\infty$ and $\varepsilon\to0$. The catalyst and shield dimensions are finite for each protocol but may grow with $n$ and the required accuracy.

\begin{figure}[htbp]
    \centering
    \includegraphics[width=0.78\textwidth]{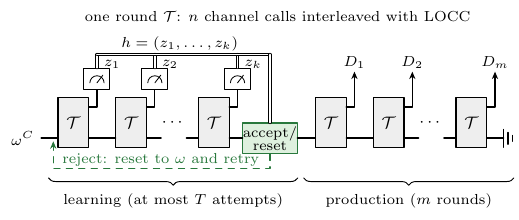}
    \caption{Learning and production for a channel protocol. Each $\mathcal T$ uses $n$ channel calls. Local IC measurements and rejection sampling follow Proposition~\ref{prop:converse}; rejected attempts reset the free product memory $\omega$. At most $T$ learning attempts of $k$ rounds each are followed by $m$ production rounds, using at most $n(m+Tk)$ channel calls in total.}
    \label{fig:channel-learning}
\end{figure}

Apply the learning proof of Proposition~\ref{prop:converse} with one invocation of $\mathcal T$ as a round, as in Fig.~\ref{fig:channel-learning}. Exact return supplies stationarity. Initialize a full-rank product memory $\omega\ge c\tau$; the local IC measurements, rejection sampling, and resets are LOCC. For every $\alpha>0$, estimate~\eqref{eq:uniform} allows learning parameters $k,T$ satisfying $K_M\eta_k+2(1-c)^T<\alpha$, independently of the production length $m$. Thus the unconditional output $\Gamma_m$, including failure branches, satisfies
\begin{equation}
    \max_i\norm{\Gamma_m^{D_i}-\Omega^D}<\alpha,
    \qquad
    \#\text{ channel calls}\le n(m+L_\alpha),
    \label{eq:channel-marginal-error}
\end{equation}
where $L_\alpha:=Tk$ is independent of $m$.

Independently repeating this entire preparation and then distilling key yields $K_D(\Gamma_m)$ per output block. Strong superadditivity and lower semicontinuity therefore give
\begin{equation}
    P_2(\mathcal N)
    \ge\frac{K_D(\Gamma_m)}{n(m+L_\alpha)}
    \ge\frac{m}{m+L_\alpha}
       \frac{K_D(\Omega^D)-o_\alpha(1)}n.
    \label{eq:channel-key}
\end{equation}
Here $o_\alpha(1)\to0$ uniformly in $m$, since all marginals approach the same fixed state $\Omega^D$. The systems $D_1,\ldots,D_m$ may be correlated; independence is needed only between repetitions of the complete preparation. Taking first $m\to\infty$ and then $\alpha\to0$ gives $P_2(\mathcal N)\ge K_D(\Omega^D)/n$.

Finally, suppose $\norm{\Omega^D-\gamma_d}\le\varepsilon$ for a private state with $d$-dimensional key registers. Key measurement followed by the Devetak--Winter bound and conditional-entropy continuity gives $K_D(\Omega^D)\ge[1-f(\varepsilon)]\log_2d$, where $f(\varepsilon)\to0$ independently of the key and shield dimensions~\cite{horodecki2009private,devetak2005,winter2016entropy}. Hence
\begin{equation}
    P_2(\mathcal N)\ge[1-f(\varepsilon)]\frac{\log_2d}{n}.
    \label{eq:channel-converse}
\end{equation}
Taking a vanishing-error sequence of catalytic protocols gives $P_2\ge P_{2,\mathrm{cc}}$. A trivial catalyst gives the reverse inequality, so $P_{2,\mathrm{cc}}(\mathcal N)=P_2(\mathcal N)$ for every finite-dimensional channel. The learning limits are taken for each fixed original protocol, so catalyst and shield dimensions may grow arbitrarily with the block length and accuracy.

\end{document}